\documentclass[slac_one]{revtex4}
\usepackage{graphicx}
\usepackage{fancyhdr}
\begin{document}

\title{Productions of Double Hypernuclei with antiprotons at PANDA} %% Paper title goes here


\author{Katarzyna Szyma\'nska }
\affiliation{Politecnico di Torino, Torino, Italy}
\affiliation{Instituto Nazionale di Fisica Nucleare, Sezione di Torino, Italy}

\begin{abstract}
One of the goals of  hypernuclear physics is to study the properties of baryon-baryon interaction including the strangeness contribution. Double hypernuclei can provide information about the  $\Lambda\Lambda$ interaction in addition to the hyperon-hyperon and hyperon-nucleus one. A new technique for producing double hypernuclei using antiprotons is foreseen in the PANDA experiment at FAIR. Gamma ray spectroscopy is a way to measure the hyperon pair binding energy and the HPGe detectors can achieve the necessary resolution.
%The HPGe detectors which achieved excellent resolution give the perfect possibility for understanding these processes. Moreover, the new production technique of  $\Lambda\Lambda$ hypernuclei  will be described.  
%
%The  preliminary results about the expected rates of stopped $\Xi^{-}$ and the resolution of the Ge crystals inside the fringing field will be presented.
\end{abstract}

\maketitle

\thispagestyle{fancy}


\section{INTRODUCTION} % Section title should be in all capitals.

The physics of the doubly strange hypernuclei, ($\Xi^{-}$) hypernuclei and double ($\Lambda\Lambda$) hypernuclei, presents some novelties with respect to the traditional hypernuclear investigations.
$\Xi^{-}$ atoms, whose formation is an intermediate step toward the $\Lambda\Lambda$-hypernucleus, can give information about the interplay of the Coulomb and strong force in the low nuclear density region, while the $\Xi^{-}$ decay inside the nucleus can shed light on the YN interaction at S=-2.

%Hypernuclear physics has several novel features in a view of baryon-baryon interactions.  
It is worth to underline that the $\Lambda\Lambda$ hypernuclei are the only possible tool to investigate the hyperon-hyperon interaction.
In fact the elementary hyperon-nucleon scattering data are scarce and the hyperon-hyperon data are totally absent. The crucial parameter in testing the potential models is the difference between the binding energy of the  $\Lambda\Lambda$ hypernucleus and twice the binding energy of each $\Lambda$ in the core nucleus. This parameters is experimentally achievable through spectroscopy measurements. Very interesting aspects of the doubly strange systems are the different contributions of the strange and not strange mesons to the $\Xi N$, $\Lambda\Lambda$, and  $\Xi N$-$\Lambda\Lambda$ coupling  interaction mechanism.
% the binding energy excess $\Delta B_{\Lambda\Lambda}$ is a crucial parameter in testing the potential models.
%However the elementary hyperon-nucleon and hyperon-hyperon scattering data are still very scarce or absent.
% The different contributions of the strange mesons to the  $\Xi N $and $\Lambda\Lambda$ interaction and to the $\Xi N$ - $\Lambda\Lambda$ coupling depend on the interaction mechanisms. 
 Moreover, the production of hyperfragments is related to the hyperon-hyperon interaction mechanism and can be a test for different models \cite{Dover,Ikeda}.  Also the $\Lambda$ decay processes show a peculiar aspect in the double hypernuclei: mesonic and non-mesonic decays can occur simultaneously and, among the non-mesonic ones, the hyperon-induced mechanism could play a role in addition to the nuclear induced one.\\

The PANDA Collaborations aims to investigate, among others, the double strangeness using the high energy antiprotons of the future FAIR facility at GSI. Measurements are planned in both nuclear spectroscopy and weak decay fields. The required high precision in spectroscopy will be achieved using Ge crystals, whose performances in the region of the fringing field have been already successfully tested. The weak decay products will be detected by Si-$\mu$strips located in a suitable arrangement together with the hypernuclear target around the antiproton beam pipe.
 %Within the future FAIR complex at GSI, the new PANDA experiment will  produce  $\Xi^{-}$‘s  using high energy antiprotons via the reaction: $\bar{p} +  N \rightarrow \Xi^{-} +  \bar{\Xi}$. 
%The major challenge at PANDA is the operation of the Ge crystal detectors close to a magnet, within the fringing field. In order to obtain precision X-ray spectroscopy of single and double hypernuclei which  extract information on  their nuclear structure.

%\section{HYPERNUCLEAR PHYSICS}

%A single hypernucleus is a nucleus that consists of protons, neutrons and one hyperon, e.g. lambda baryon.
%The physics of the double strange system ( $\Lambda\Lambda$ hypernuclei and $\Xi^{-}$ hypernuclei) presents several appealing aspects.
%First of all, it must be remarked that the $\Lambda\Lambda$ hypernuclei are the only possible tool to investigate the hyperon-hyperon interaction: the binding energy excess  $\Delta B_{\Lambda\Lambda}$ is a crucial parameter in testing the potential models, while the different contributions of the strange mesons to the  $\Xi N $and $\Lambda\Lambda$ interaction and to the $\Xi N$ - $\Lambda\Lambda$ coupling depend on the interaction mechanisms. Moreover, the hyperfragments production is related to the hyperon-hyperon interaction mechanism and can be a 
%test for different models.  Also the $\Lambda$ decay processes show a peculiar aspect in the double hypernuclei: mesonic and non-mesonic decays can occur simultaneously and, among the non-mesonic ones, the hyperon-induced mechanism could play a role in addition to the nuclear induced one. 

\section{PRODUCTION OF HYPERNUCLEI AT PANDA } 

The FAIR (Facility for Antiproton and Ion Research) complex  \cite{Spiller,Henning}, located in the GSI site, will include the ring HESR (High Energy Storage Ring)  to store antiprotons of energy between 0.8 to 14.4 MeV. Intense and high quality beams are foreseen:
luminosity up to $10^{32}cm^{-2}s^{-1}$ and momentum resolution up to $10^{-5}$ are expected.
 %project, the storage ring HESR \cite{Lehrac} will supply intense antiproton beams of high quality in the momentum range from 1.5 to 15 GeV/c. HESR   will host the PANDA experiment, dedicated to the hadronic physics, mainly charm and hypernuclear physics.  
%The experimental set-up of PANDA consist of an assembly of two magnets and several detectors. It is designed to surround the antiproton line of HESR in a cylindrical symmetry. Since the threshold to produce all strange and charmed baryon-antibaryon pairs goes from 1.44 Gev/c to 14.6 GeV/c, the production of  $\Xi^{-} \bar{\Xi}$ is fully accessible as well as the other pairs.

The new technique  proposed by the PANDA Collaboration \cite{Pochodzalla} to  produce double hypernuclei with antiprotons, is based on the reaction:
% $\Lambda\Lambda$ pairs uses antiprotons at 3GeV/c for the reaction

%The new technique \cite{Pochodzalla} proposed by the PANDA Collaboration to  produce  $\Lambda\Lambda$  using antiprotons with momentum of 3 GeV/c via the reaction
\begin{equation}
\bar{p} +  N \rightarrow \Xi^{-} + \bar{\Xi}
\end{equation}    
In this reaction two different targets are used, the first one (primary target) where $\Xi^{-}$ is produced and the second one (secondary target) where $\Lambda\Lambda$ hypernucleus is formed. 

%Two targets are foreseen: in the firs (primary) target the $\Xi^{-}$ is produced, in the second the $\Lambda\Lambda$ hypernucleus is formed.
 The whole process proceeds through the following steps:
a) reaction (1) occurs quasi free in a nucleus,
b) $\Xi^{-}$ re-scatters in the residual nucleus and it is strongly decelerated,
c) $\Xi^{-}$ slows down to stop,
d) is captured by an atom,
e) undergoes to an atomic cascade,
f) is captured into the nucleus,
g) makes the conversion reaction:

%a) re-scattering of  $\Xi^{-}$ in the residual nucleus, in a primary target and slowing down to stop in a secondary target,
%b) atomic capture and decay, capture in the nucleus of a secondary target and decay, 
%c) conversion reaction
\begin{equation}
\label{conversion}
\Xi^{-} + p \rightarrow \Lambda\Lambda 
\end{equation}

Steps a) and b) occur in the primary target while the other ones in the secondary. After reaction (2) both $\Lambda'$s can stick to the nucleus or not: the fact that the excess of energy is 28 MeV makes the probability of both sticking rather high.
%One of the most important feature for producing $\Lambda\Lambda$ in reaction( \ref{conversion}) is the fact that of excess energy is 28MeV, thus making  the probability of both $\Lambda$ to stick to the nucleus rather high.

\begin{figure}[!htbp]
   
   \begin{center}
         \includegraphics[width=7.5cm]{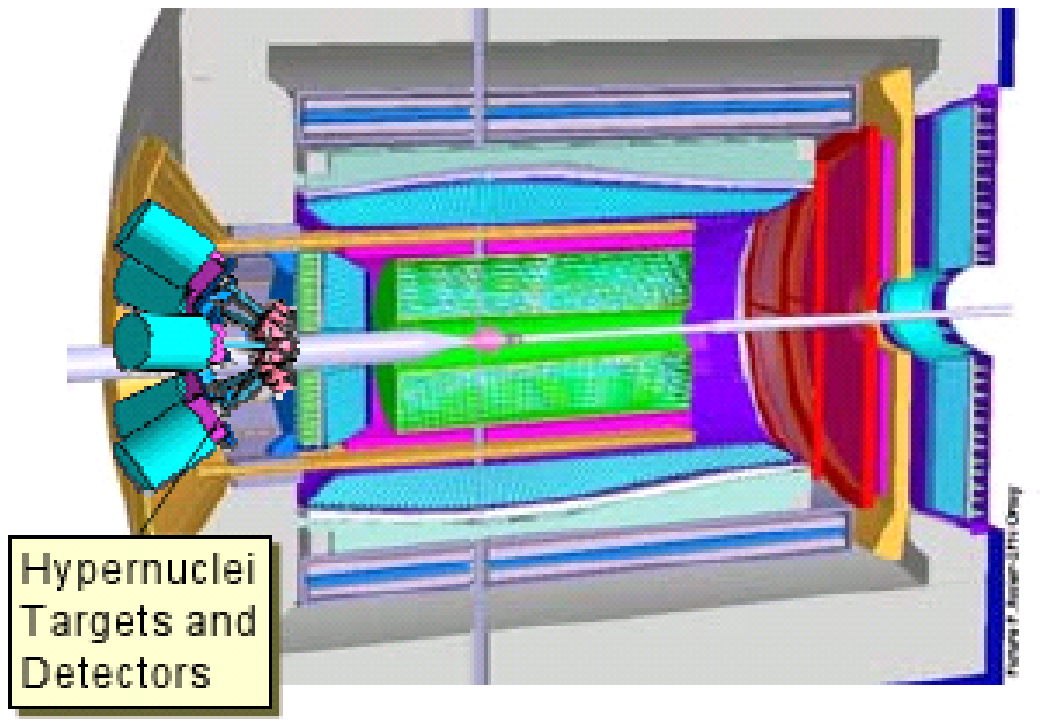}
     \includegraphics[width=10cm]{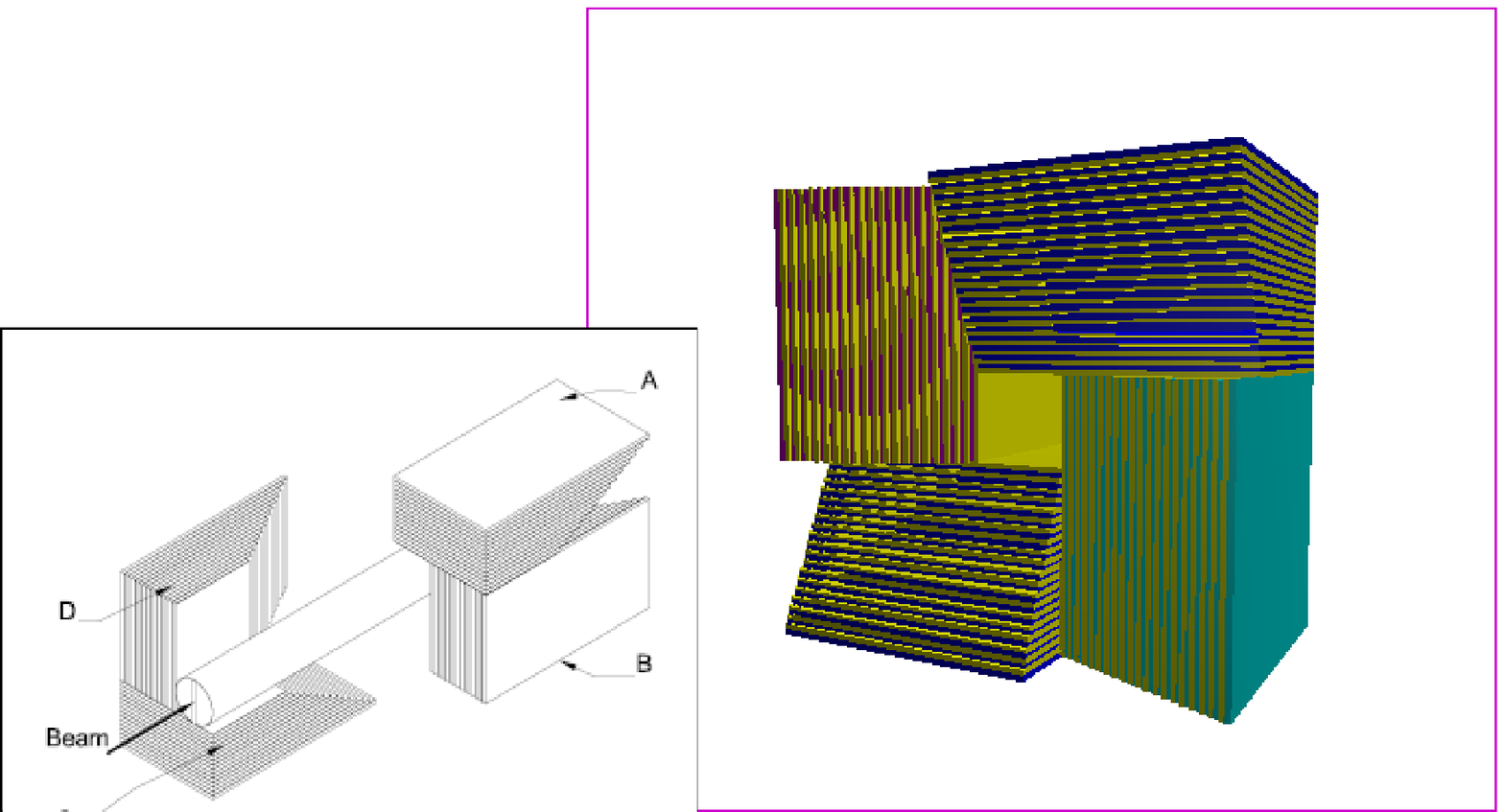} 
     
     \caption{{\em Scheme of PANDA detector system with the assembled of Ge crystal \cite{fair} (left side) and the target system for double hypernuclei production, on the right.}}
 \label{panda}
   \end{center} 
 \end{figure}

The designed  apparatus for Double Hypernuclei physics is sketched in Fig.\ref{panda} (left) and includes the primary and secondary targets in expanded view in Fig.\ref{panda} (right).
The efficiency of this target arrangement has been evaluated by simulation \cite{Ferro} and the ratio R of the stopped $\Xi^{-}$ in the secondary target to the  number of produced $\Xi^{-}$ in reaction (1) is reported in Table \ref{rates} as a function of  different primary targets for a fixed ${}^{12}C$ secondary target.

% which include primary and secondary targets, is sketched  in Fig.\ref{panda}( right side). 
%The primary target is designed as a thin wire (20$\mu$m diameter ) of ${}^{12}C$ located inside a beam pipe, perpendicular to the beam direction: materials and sizes are  chosen to optimize the $\Xi^{-}$ production and the slowing down to stop. The secondary target,
% is playing a crucial role for stopping $\Xi^{-}$ in an ordinary matter.
%is designed  as 4 blocks located around the beam pipe, each one with a compact sandwich-structure of alternate target layers and Si-$\mu$strips used to detect the weak decay products.
%detectors used for absorption of low energy hyperons at large angle and in order to tracks pions and protons  from the hypernucleus decay.
% of mixed layers of passive and active (Si-$\mu$strips) material in order to detect pions and protons  from the hypernucleus decay.
% The expected rates of the $\Xi^{-}$ stopped in a ${}^{12}C$ secondary target calculated in the Intra-Nuclear Cascade model  vary from 0.2\% for the light material  up to 0.4 \% for the heavier one \cite{Ferro}.
%Moreover, beside the S = -2 hyperon $\Xi^-$, an anti-hyperon is released too. This antiparticle is playing a crucial role in the experimental detection of the whole double hypernuclei formation because  it may be also used as trigger of the $\Xi^{-}$ productions.

The efficiency depends slightly on the mass number of the primary target material: for secondary targets of heavier nuclei  one should expect an improvement of the efficiency, due to a faster slowing down to stop of $\Xi^{-}$.
Therefore an efficiency of the order of $10^{-3}$ has to be combined with the rate of $\Xi^{-}$ produced at HESR, to get the $\Xi^{-}$ stopped rate. The production rate will depend on the beam profile and on the constraints due to the radiation damage of the PANDA detectors. The designed structure of the antiproton beam \cite{Lehrac} and a preliminary estimate of the radiation tolerance of $\approx$ $6\cdot10^{6}$ charged particle/s allow to expect a rate of $\approx$ 0.17 stopped $\Xi^{-}$/s using a ${}^{12}C$ primary target. Such a rate will produce in one month a number of stopped $\Xi^{-}$ higher than the existing statistics.
Concerning the efficiency in detecting the $\Xi^{-}$, it must be remarked that also an anti-hyperon is released in reaction (1), which annihilates mostly inside the primary target. This annihilation produces (at least) two anti-kaons and this peculiar aspect will be used as a  tool for a high-level trigger.

\subsection{X and gamma rays detection with HPGe crystals}

%The double hypernuclear spectroscopy is the only way to explore the Y−Y interaction from the experimental side.
%Atomic cascade, nuclear cascade and conversion reaction produce X and gamma-rays which give information about the binding energy, if high resolution detectors are used.

Detection of X-rays from hyperatoms \cite{Batty, Gal} and gamma-rays from double hypernuclear processes plays a crucial role in the hypernuclear program of PANDA. To achieve the required resolution, High Purity Germanium detectors (HPGe) are planned to be installed.

% The low energy pions and protons coming from the decay of the hyperons and hypernuclei, detected by a tracker close to the secondary target, can give information about the weak mesonic and non mesonic decay of the hyperons.

% in future PANDA hypernuclei experiments.  In PANDA due to the technique for the strangeness production based on the antiproton-nucleus  interaction \cite{Pochodzalla2} not only the double hypernuclei but also the doubly strange atoms \cite{Gal,Batty} could be explored.
A common feature of the general purpose apparatuses, like PANDA, is the presence of large spectrometers with intense magnetic fields. In the PANDA set-up a set of HPGe detectors will be located upstream the target (see Fig.\ref{panda}, left), just in front of the upstream end of the solenoid. The set of Ge crystals will be assembled to cover almost 2$\pi$ solid angle and they will be located in the area when the the fringing field is not negligible. 
Due to the magnetic field effects on the charge carrier trajectories the performances of the Ge detectors could be modified as well as the electronic circuits could be affected. Moreover the data acquisition lasts in general several weeks or months. Therefore it is mandatory to check whether relevant changes in the performances of the HPGe detector might occur and if such changes are depending on the time spent in operating inside the magnetic field. Tests of long duration (about 1 year) have been performed on a HPGe crystal of the same sizes and features of those ones planned for PANDA  \cite{Szymanska}. The result showed that the worsening of the resolution was within 10\% in the $\gamma$ range from 0.08 MeV to 1.333 MeV and that such a worsening was not at all permanent. These results allow to design the relative position of the crystals and of the primary target in order to maximize the acceptance.

% Study and result of properties for HPGe crystals inside magnetic field are described in \cite{Szymanska}.  Moreover the data acquisition lasts in general several weeks or months.
%The aim of that work was to evaluate the changes in resolution, efficiency and “tailing” of the crystal, due to the field at different energies. The scanned energy range was from 0.08 MeV to 1.333 MeV.

 %The X rays from the $\Xi^-$-atom and $\gamma$'s from the $\Xi^-$ -and $\Lambda\Lambda$ -hypernuclear states will be detected by a set of HPGe crystals located upstream the target (see Fig.\ref{panda}, left side). The set of Ge detectors will be assembled to cover almost 2 $\pi$ solid angle.  Ge detectors can provide the sufficient precision for measuring levels, shifts and widths of the doubly strange systems.

\begin{table}[t]
\begin{center}
\caption{Rates of the stopped $\Xi^{-}$ inside the secondary target to the produced $\Xi^{-}$ in the primary target via reaction (1).}

\begin{tabular}{|c|c|c|c|c|c|c|}
\hline \textbf{Primary target} & \textbf{${}^{12}C$} &\textbf{${}^{72}Al$}&\textbf{${}^{59}Ni$}&\textbf{${}^{108}Ag$}&\textbf{${}^{137}Ba$}&\textbf{${}^{177}Au$}
\\
\hline Rate (\%) & 0.213 & 0.268 & 0.325&0.352&0.360&0.391 \\
\hline
\end{tabular}
\label{rates}
\end{center}
\end{table}

 \section{CONCLUSIONS} 
 
 A new technique for producing $\Xi^{-}$ and double hypernuclei from antiproton at 3 GeV/c  has been proposed in PANDA experiment. The designed two targets system will allow different hypernuclei be formed with good efficiency. The expected rates of stopped $\Xi^{-}$ have been estimated to be sufficient in increasing the present amount  of the double hypernuclei data up to a statistically significant amount.
High resolution and stability of the HPGe in detecting the X and $\gamma$ rays will be of basic importance in the measurements of the hypernuclear system.
%The use of the crystals close to the spectrometer requires a check whether their performances will be maintained inside a fringing field.
% and high amount of data is expected.
 Preliminary tests of the detectors inside the magnetic field have been successfully performed.\\
Both high $\Xi^{-}$ rates and performances are promising a step forward in understanding many important aspects of the strong interaction in the S=-2 systems and in the hyperon weak decay.
%Precision spectroscopy with high resolution Ge detectors give the possibility to understand  many aspects of the strong  interaction in the S=-2 systems and the hyperon weak decay.

% A lot of works are of course still necessary in order to fully design the geometrical arrangement, the sizes and the materials of the targets, in order to optimize the production and to meet the beam requirement of the

\end{document}